\documentclass[aps,prl,twocolumn,showpacs,superscriptaddress]{revtex4-2}
\usepackage{graphicx}
\usepackage{amsmath}
\usepackage{multirow}
\usepackage{tabularx}
\usepackage{color}
\usepackage[colorlinks,linkcolor=blue,urlcolor=blue,anchorcolor=blue,citecolor=blue]{hyperref}
\begin{document}
\preprint{APS/123-QED}
\title{Bilayer two-orbital model of La\textsubscript{3}Ni\textsubscript{2}O\textsubscript{7}
under pressure}
\date{\today}

\author{Zhihui Luo}
\thanks{These authors contributed equally to this work}
\author{Xunwu Hu}
\thanks{These authors contributed equally to this work}
\author{Meng Wang}
\author{W\'ei W\'u}
\author{Dao-Xin Yao}
\email{yaodaox@mail.sysu.edu.cn}
\affiliation{Center for Neutron Science and Technology, Guangdong Provincial Key Laboratory of Magnetoelectric Physics and Devices, State Key Laboratory of Optoelectronic Materials and Technologies, School of Physics, Sun Yat-Sen University, Guangzhou, 510275, China}

\begin{abstract}
The newly discovered Ruddlesden-Popper bilayer La$_3$Ni$_2$O$_7$ reaches an remarkable superconducting transition temperature $T_c=80$\ K  under a pressure of above 14\ GPa. Here we propose a minimal bilayer two-orbital model of the high-pressure phase of La$_3$Ni$_2$O$_7$. Our model is constructed with the Ni$-3d_{x^{2}-y^{2}}$, $3d_{3z^{2}-r^{2}}$ orbitals by using Wannier downfolding of the density
functional theory calculations, which captures the key ingredients of the material, such as band structure and Fermi surface topology. There are two  
electron pockets $\alpha,\beta$ and one hole pocket $\gamma$ on the Fermi surface, in which the $\alpha,\beta$ pockets show mixing of two orbitals, while the  $\gamma-$pocket is associated with Ni$-d_{3z^{2}-r^{2}}$  orbital. The random phase approximation spin susceptibility reveals a magnetic enhancement associating to the $d_{3z^2-r^2}$ state. A higher energy model with O$-p$ orbitals is also provided for further study.
\end{abstract}
\maketitle


{\it Introduction.}$-$Recently the newly discovered Ruddlesden-Popper bilayer perovskite
nickelate La\textsubscript{3}Ni\textsubscript{2}O\textsubscript{7}
shows a remarkable high superconducting transition temperature of
$T_c=80$ K with an applied pressure of over 14\ GPa \cite{sun2023}. This breakthrough will undoubtedly provoke a stir in the field of high$-T_c$ superconductivity long after the discovery of cuprate \cite{bed1986,anderson1987,lee2006,keimer2015,saka2014,liw2019} and iron-based \cite{raghu2008,chubu2008,esch2009,dag2010}, as well as the recent infinite layer Nickelate superconductors \cite{li2019,been2021,lee2004,hepting2020,botana2020,wu2020,jinag2020,hu2019,Nomura2019,zhangya2020,werner2020,berna2020,gu2020,kitatani2020nickelate,lech2021,kre2022,lech2020,kita2020,zhang2020}.  At ambient pressure, La\textsubscript{3}Ni\textsubscript{2}O\textsubscript{7}
exhibits an orthorhombic structure of Amam space group \cite{liu2023}. With increasing
pressure, it undergoes a structure transition to Fmmm space group, which possesses a more regular AA-stacking structure
with apical Ni-O-Ni bond approaching 180$^{\circ}$ [see Figs.\ 1f, g in Ref.~\cite{sun2023}]. 
The most crucial effect of the pressure is to drive a metallic transition of the correlated electronic ground state. 
The resistance measurement shows that, above $T_c$, La$_3$Ni$_2$O$_7$ undergoes a transition from weakly insulating  to metallic phase [see Figs.\ 3a, 4 in Ref.~\cite{sun2023}], 
which is further evidenced in the density functional theory (DFT) calculation as the emergence of an additional Ni$-d_{3z^2-r^2}$ state near Fermi energy ($E_\mathrm F$) \cite{sun2023,pardo2011}. Such a state is essentially associated to the $\sigma-$bonding that connects Ni$-d_{3z^2-r^2}$ and apical O$-p_z$ orbitals, further indicating a rather different situation in La$_3$Ni$_2$O$_7$ in which the unconventional pairing might be promoted by such a coupling degree \cite{choi2002,gao2015,dro2019,gao22015,hu2016i,hu2015prx}. Therefore, it is of vital at once to understand the effective low-energy physics.

In this paper, we propose a bilayer two-orbital model for the high-pressure phase of La\textsubscript{3}Ni\textsubscript{2}O\textsubscript{7}. Our models is constructed based on Wannier downfolding of the DFT band structure,
which capture key feature of electronic structure at $E_\mathrm{F}$ and could serve as a starting for further strongly correlated calculations and investigation on the unconventional pairing symmetry. 

\begin{figure}
\includegraphics[scale=0.45,trim=24 0 0 0]{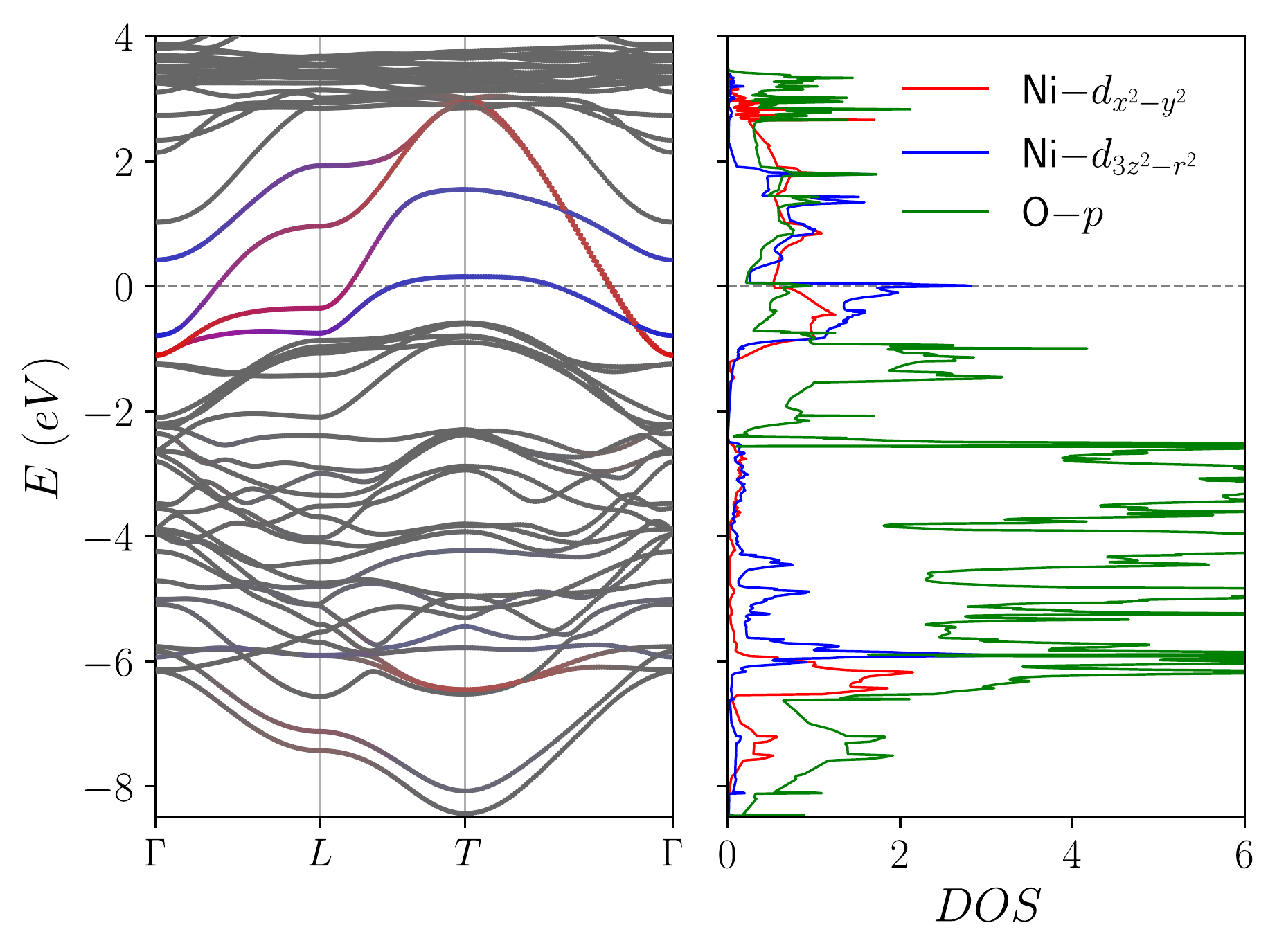}\caption{\label{fig:dft}The DFT band structure and partial density of states of the high-pressure Fmmm phase of La\textsubscript{3}Ni\textsubscript{2}O\textsubscript{7}. The blue, red and green colors represent Ni$-d_{x^2-y^2}$, $d_{3z^2-r^2}$ and O$-p$ states, respectively.}
\end{figure}

{\it Electronic model.}$-$  To elucidate the electronic structure of La\textsubscript{3}Ni\textsubscript{2}O\textsubscript{7} under the high-pressure phase (29.5\ Gpa), a primitive unit cell with two-Ni atoms is adopted. We fix the experimentally refined lattice parameters \cite{sun2023} and fully optimize the atomic positions using the DFT as implemented in the Vienna {\it ab initio} simulation package (VASP) \cite{93prb,96prb}. The projector augmented-wave (PAW) method \cite{94prb} with a 600 eV plane-wave cutoff is adopted. The generalized gradient approximation (GGA) of Perdew-Burke-Ernzerhof (PBE) \cite{96prl} is used for exchange-correlation functional.
In Fig.~\ref{fig:dft} we show the resulting band structure and partial density of states, which distinctly shows a major Ni$-d_{x^2-y^2}$ and $d_{3z^2-r^2}$ near $E_\mathrm{F}$. Note that there appears a hole pocket associating Ni$-d_{3z^2-r^2}$ orbital at $T$ point. This hole pocket is separated from the other one upper the Fermi surface with an energy $\sim 1.3$\ eV. This splitting should be attributed to $pd\sigma-$bonding between Ni$-d_{3z^2-r^2}$ and apical O$-p_z$ orbitals.

\begin{figure}
\includegraphics[scale=0.35,trim=10 0 0 0]{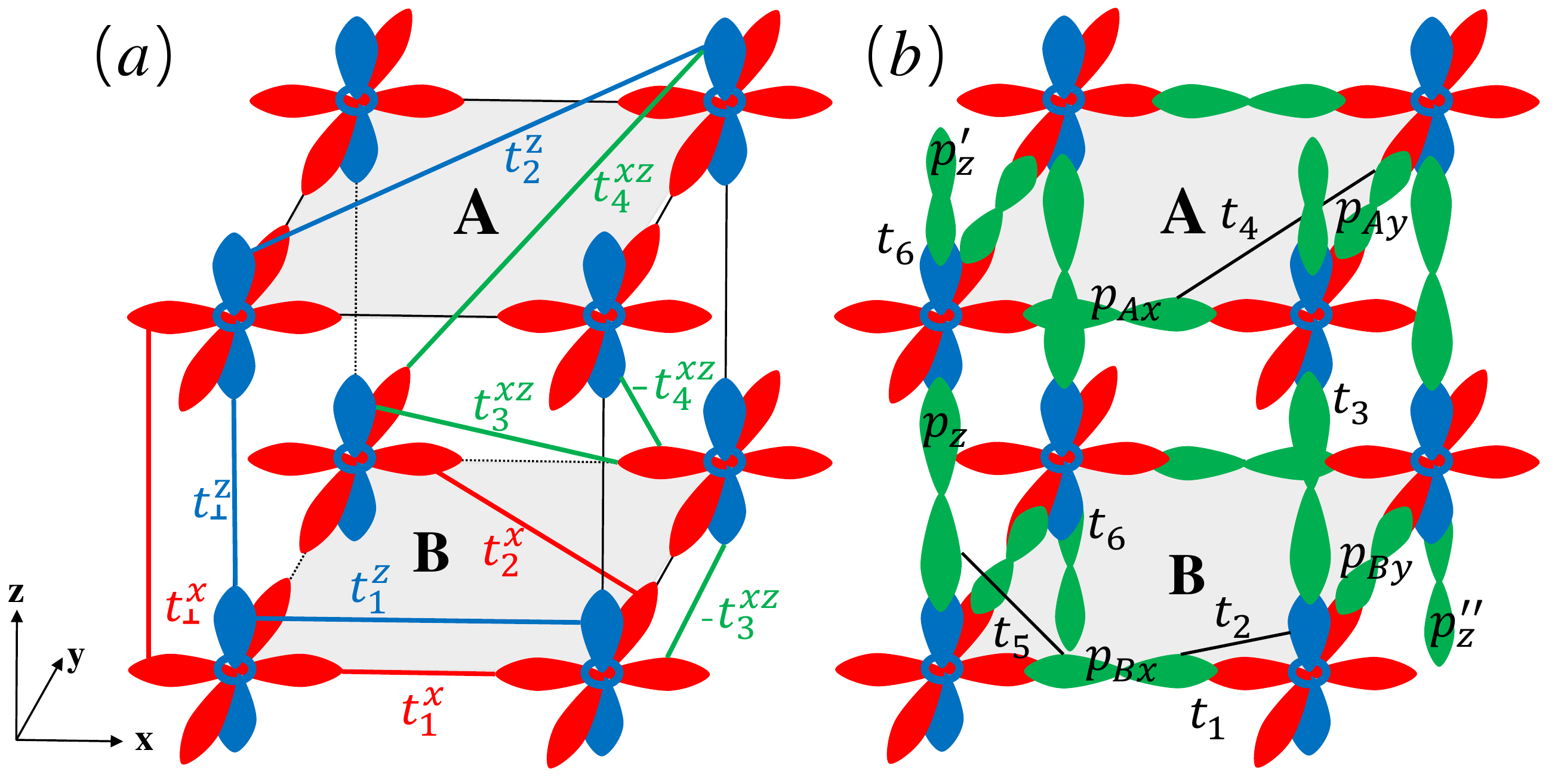}\caption{\label{fig:1}Schematic of the bilayer La$_3$Ni$_2$O$_7$ lattice with hopping parameters.  (a) Only Ni$-d_{x^{2}-y^{2}}$ (red), $d_{3z^{2}-r^{2}}$ (blue) orbitals are shown. The blue, red, green lines indicate hoppings for the bilayer two-orbital model.  Their values are listed in Tabel~\ref{tab:hop1}. (b) Extra  O$-p$ orbitals are drawn as green shapes, with in-plane $p_x,p_y$ and apical $p_z,p_z^\prime,p_z^{\prime\prime}$. Some of the $p_z^\prime,p_z^{\prime\prime}$ are hidden for clarity. The hopping parameters are given in Tabel~\ref{tab:hop2}.}

\end{figure}

With the Wannier downfolding \cite{2020jpcm,97prb,2001prb} of the DFT band structure, we arrive at an effective bilayer two-orbital model


\begin{align}
\label{eq:H} 
\mathcal{H}&= \mathcal{H}_{0}+\mathcal{H}_U,  \\
\mathcal{H}_{0}&=\sum_{{\rm k}\sigma}\Psi_{{\rm k}\sigma}^{\dagger}H({\rm k})\Psi_{{\rm k}\sigma}, \nonumber  \\
\mathcal{H}_{U}&=U\sum_{is}n_{is\uparrow}n_{is\downarrow} \nonumber\\
&+\sum_{i\alpha \beta}(U^\prime-J\delta_{\alpha \beta})(n_{iAx\alpha}n_{iAz\beta}+n_{iBx\alpha}n_{iBz\beta}). \nonumber
\end{align}
Here $\mathcal{H}_0$ is the tight-binding Hamiltonian determined out of our Wannier downfolding, and $\mathcal{H}_U$ is the Coulomb interaction term \cite{kanamori}. The basis is defined as $\Psi_{\sigma}=\left(d_{Ax\sigma},d_{Az\sigma},d_{Bx\sigma},d_{Bz\sigma}\right)^{T}$, 
with the field operator $d_{s\sigma}$ 
denotes annihilation of an $s=Ax,Az,Bx,Bz$ 
electron with spin $\sigma$. As shown in Fig.~\ref{fig:1}, $A,B$ label the bilayer, and $x,z$ label $d_{x^2-y^2},d_{3z^2-r^2}$ orbitals, respectively. For $\mathcal{H}_U$, 
$U,U^\prime,J$ are intra-orbital, inter-orbital Coulomb repulsion and Hund coupling, respectively.
The matrix $H(\mathrm{k})$ is written as
\begin{align}
H & ({\rm k})=\left(\begin{array}{cc}
H_{A}({\rm k}) & H_{AB}({\rm k})\\
H_{AB}({\rm k}) & H_{A}({\rm k})
\end{array}\right),\nonumber \\
H_{A}({\rm k})= & \left(\begin{array}{cc}
T_{{\rm k}}^{x} & V_{{\rm k}}\\
V_{{\rm k}} & T_{{\rm k}}^{z}
\end{array}\right),\qquad H_{AB}({\rm k})=\left(\begin{array}{cc}
t_{\bot}^{x} & V_{{\rm k}}^{\prime}\\
V_{{\rm k}}^{\prime} & t_{\bot}^{z}
\end{array}\right).\label{eq:tb}
\end{align}
with
\begin{align*}
T_{{\rm k}}^{x/z} & =2t_{1}^{x/z}\left(\cos  {k}_{x}+\cos {k}_{y}\right)+4t_{2}^{x/z}\cos {k}_{x}\cos {k}_{y}+\epsilon^{x/z},\\
V_{\rm{ k}} & =2t_{3}^{xz}\left(\cos {k}_{x}-\cos {k}_{y}\right),\quad V_{\rm{ k}}^{\prime}=2t_{4}^{xz}\left(\cos {k}_{x}-\cos {k}_{y}\right).
\end{align*}
Here  $T^{x/z}_{\rm k}$ represents intra-layer intra-orbital hopping, and $V_{{\rm k}}$
($V_{{\rm k}}^{\prime}$) represent intra-layer (inter-layer) hybridization
between $d_{x^{2}-y^{2}}$ and $d_{3z^{2}-r^{2}}$ orbitals.
The essential hoppings $t_1^{x/z},t_2^{x/z},t_{3}^{xz},t_{4}^{xz}$ are demonstrated in Fig.~\ref{fig:1}a.
Note that the minus sign appeared in the structure factor of $t_{3}^{xz},t_{4}^{xz}$ is associated to the orbital symmetry of two $e_g$ sectors. 



\begin{table}
\begin{tabular}{ccccc}
\hline \hline 

$t_{1}^{x}$ & $t_{1}^{z}$ & $t_{2}^{x}$ & $t_{2}^{z}$ & $t_{3}^{xz}$  \tabularnewline\hline 
-0.483 & -0.110 & 0.069 & -0.017 & 0.239  \tabularnewline \hline 

 $t_{\bot}^{x}$ & $t_{\bot}^{z}$ & $t_{4}^{xz}$ & $\epsilon^{x}$ & $\epsilon^{z}$  \tabularnewline \hline 
 0.005 & -0.635 & -0.034 &  0.776 & 0.409 \tabularnewline
 
 \hline \hline
\end{tabular}
\caption{\label{tab:hop1}Tight-binding  parameters  of the bilayer two-orbital model. The hoppings $t$ are demonstrated in Fig.~\ref{fig:1}a.  $\epsilon^{x},\epsilon^z$ are site energies for Ni$-d_{x^2-y^2}, d_{3z^2-r^2}$ orbitals, respectively.    
}
\end{table}


To better illustrate the low-energy state, it is advisable to further simplify the above model. Recall that the mirror symmetry of the bilayer structure allows us to define the bonding and anti-bonding states $\Phi_{\pm{\rm k}\sigma}=\left(c_{\pm{\rm k}\sigma}^{x},c_{\pm{\rm k}\sigma}^{z}\right)^{T}$
with $c_{\pm{\rm k}\sigma}^{x/z}=\frac{1}{\sqrt{2}}\left(d_{{\rm k}A\sigma}^{x/z}\pm d_{{\rm k}B\sigma}^{x/z}\right)$,
 in which the Hamiltonian acquires a block-diagonal form
\begin{align}
\mathcal{H}_{0} & =\sum_{{\rm k}\sigma}\left(\Phi_{+{\rm k}\sigma}^{\dagger}H_{+}({\rm k})\Phi_{+{\rm k}\sigma}+\Phi_{-{\rm k}\sigma}^{\dagger}H_{-}({\rm k})\Phi_{-{\rm k}\sigma}\right),\nonumber \\
H_{\pm}({\rm k}) & =\left(\begin{array}{cc}
T_{{\rm k}}^{x}\pm t_{\bot}^{x} & V_{{\rm k}}\pm V_{{\rm k}}^{\prime}\\
V_{{\rm k}}\pm V_{{\rm k}}^{\prime} & T_{{\rm k}}^{z}\pm t_{\bot}^{z}
\end{array}\right).\label{eq:bonding-antibonding}
\end{align}
In this representation, the two $d_{3z^2-r^2}$ states at $E_\mathrm{F}$ are manifested as the component  $T^z_\mathrm{k}\pm t_\bot^z$ which define a splitting energy $2t_\bot^z$. 

\begin{figure}
\includegraphics[scale=0.6,trim=15 0 0 0]{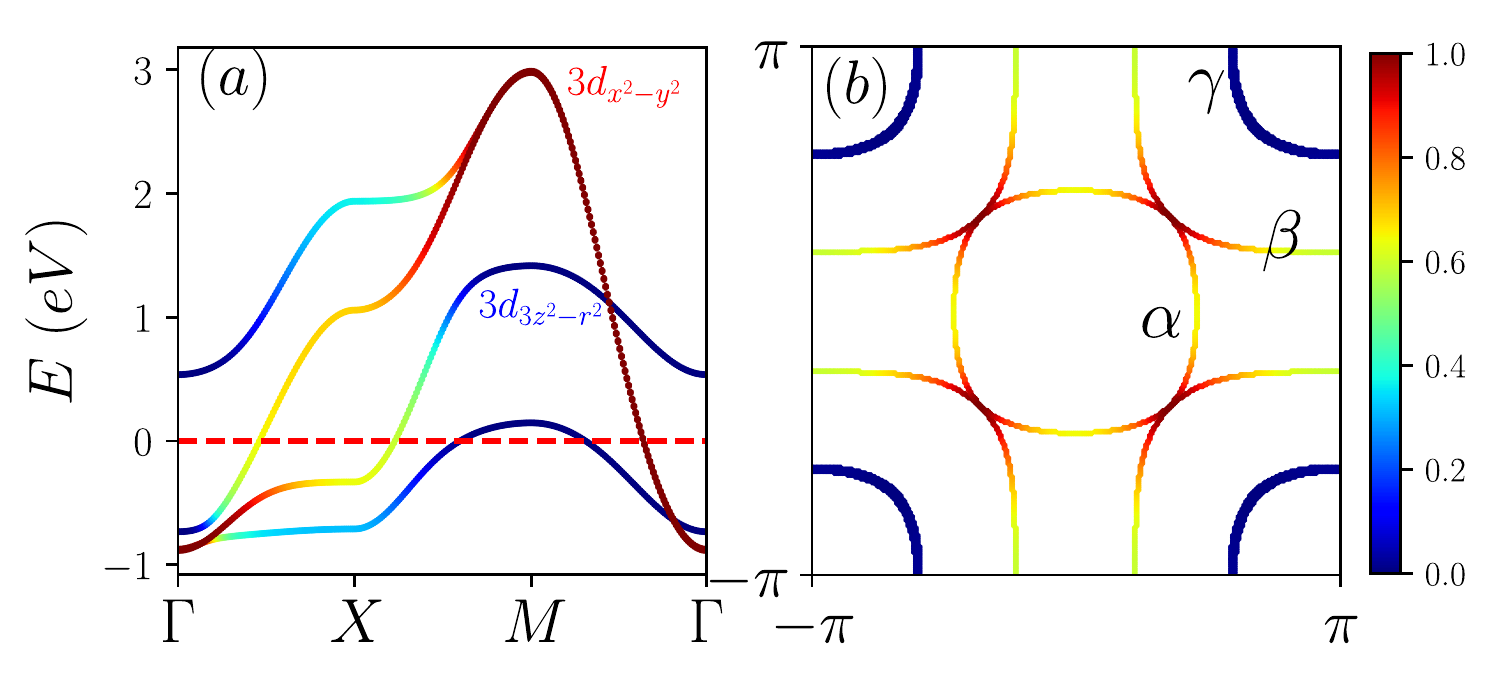}
\caption{The band structure (a) and Fermi surface (b) of the bilayer two-orbital model. The colorbar indicates the orbital weight of $d_{x^{2}-y^{2}}$ and $d_{3z^{2}-r^{2}}$.}
\label{fig:band}
\end{figure}

With the value of tight-binding parameters listed in Tabel~\ref{tab:hop1}, we show in Fig.~\ref{fig:band}  the resulting band structure and Fermi surface. The model reproduces the DFT band structure well at $E_\mathrm{F}$. 
Also, site energies are slightly adjusted to  coincide with the nominal $d^{7.5}$ configuration \cite{sun2023,pardo2011}. 
In Fig.~\ref{fig:band}b we can see two electron pockets $\alpha,\beta$ and one hole pocket $\gamma$.  The $\alpha,\beta-$pocket show  mixing of orbital content, while the  $\gamma-$pocket is featured as a dominated $d_{3z^{2}-r^{2}}$ state. 
Note that the amplitude of $t_\bot^z=-0.635$ is even larger than that of the intra-layer nearest-neighbor hopping  $t_1^x=-0.483$, by a ratio of  1.3.  
This strong inter-layer coupling indicates a  possible different situation of the unconventional paring as compared to cuprates, and is in reminiscent of a theoretical bilayer-Hubbard model \cite{maier11,Nakata2017}, in which an $s_\pm-$wave pairing could be promoted via inter-layer coupling.
But there is a key difference here. In La$_3$Ni$_2$O$_7$, $t_\bot^z$ only appears in $d_{3z^2-r^2}$ sector, while  for $d_{x^2-y^2}$ the amplitude  $t_\bot^x=0.005$ is marginal. Hence, the influence from inter-layer coupling to NiO$_2$ plane can only be achieved via hybridizations $V_\mathrm{k},V_\mathrm{k}^\prime$ . It would be interesting to see how pairing symmetry is affected in this situation.
We would also like to point out that, however, due to the  asymmetry of orthorhombic structure of this compound, the $\gamma-$pocket from DFT is  slightly  stretched along nodal direction. 




To explicitly consider the physics of O$-p$ orbitals, we introduce a higher energy model (eleven-orbital model). The basis is $\Psi=(d_{Ax},d_{Az},d_{Bx},d_{Bz},p_{Ax},p_{Ay},p_{Bx},p_{By},p_{z},p^\prime_{z},p^{\prime\prime}_{z})^T$, with four more in-plane $p_{Ax},p_{Ay},p_{Bx},p_{By}$ and three apical $p_z,p_z^\prime,p_z^{\prime\prime}$ as shown in Fig.~\ref{fig:1}b. The tight-binding parameters of the model are listed in Tabel~\ref{tab:hop2}, which requires six hopping parameters including necessary $pd,pp$ overlaps.
The resulting band structure covers an energy range akin to that of  Fig.~\ref{fig:dft} and can also reproduced the main features at $E_\mathrm{F}$. 
Moreover, we found a strong hopping of   $t_6=1.366$ between $d_{3z^2-r^2}$ and two apical $p_z^\prime,p_z^{\prime\prime}$ that lie outside the bilayer, which manifest as two hole baths for NiO$_2$ plane and could be further integrated out in a L\"owdin downfolding technique \cite{downfold04}.
The model will be useful for further study of the electronic correlation in the dynamic mean field theory framework.

 \begin{table}
\begin{tabular}{cccccc}
\hline \hline
$t_1$ & $t_2$ & $t_3$ & $t_4$ & $t_5$ & $t_6$\tabularnewline
\hline
-1.564 & 0.747 & -1.625 & 0.577  & -0.487 & 1.366\tabularnewline
\hline  
$\epsilon^{x}$ & $\epsilon^z$ & $\epsilon_{p}^{x/y}$  & $\epsilon_{p}^z$  & $\epsilon_{p\prime/p\prime\prime}^z$\tabularnewline
\hline  
-1.057 & -1.161 &-4.936  & -4.294 & -3.772 &\tabularnewline
\hline \hline
\end{tabular}
\caption{\label{tab:hop2}Tight-binding parameters for Wannier downfolding of the
eleven-orbital model. $\epsilon^{x/z}$ are site energies for  $d_{x^2-y^2}/d_{3z^2-r^2}$, and $\epsilon^{x/y}_p$ for in-plane $p_x/p_y$,  and $\epsilon^z_{p/p^\prime/p{\prime\prime}}$  for apical $p_z/p_z^\prime/p_z^{\prime\prime}$. See Fig.~\ref{fig:1}b for details.  }
\end{table}

{\it Spin susceptibility.}$-$To determine the magnetic response of the material,
we investigate the the spin susceptibility of our model, which is defined as
\begin{align}
\chi_{S}^{st}(q,i\omega_{n})=\frac{1}{3}\int_{0}^{\beta}d\tau e^{i\omega_{n}\tau}\langle\bm{S}_{s}(q,\tau)\cdot\bm{S}_{t}(-q,0)\rangle.
\label{eq:chi}
\end{align}
Here $s,t=Ax,Az,Bx,Bz$ label orbitals,
and the spin operator is defined as $\bm{S}_{qs}=\frac{1}{2}\sum_{\rm {k}\alpha\beta}d_{\rm {k}s\alpha}^{\dagger}\boldsymbol{\sigma}_{\alpha\beta}d_{\rm {k}+qs\beta}$ with $\bm{\sigma}$ the Pauli matrix.
By using wick theorem, we expand Eq. \ref{eq:chi} to obtain  the bare
(non-interacting) susceptibility

\begin{align*}
\chi_{S}^{st}(q,i\omega_{n}) & =-\frac{1}{2N}\sum_{mn}\frac{f(\epsilon_{\rm k}^{n})-f(\epsilon_{\rm k}^{m})}{i\omega_{n}+\epsilon_{\rm k}^{n}-\epsilon_{\rm {k}+q}^{m}}\\
 & \quad\times\langle m|\rm {k}+qt\rangle\langle \rm {k}+qs|m\rangle\langle n|\rm {k}s\rangle\langle \rm{k}t|n\rangle,
\end{align*}

with $m,n$ the band indices and $f(\epsilon)=\frac{1}{e^{\epsilon/T}+1}$
the Fermi-Dirac function. $\langle \rm {k}s|m\rangle$ represents the eigenvector
relating $s,m$ states at wave vector $\rm k$. 

Under the random phase approximation, the spin susceptibility is calculated by
\begin{align}
\chi^{st,\mathrm{RPA}}_{S}(q,i\omega_n)=[I-\chi_S^{st}(q,i\omega_n)\Gamma]^{-1}\chi_S^{st}(q,i\omega_n),
\end{align}
with interaction vertex defined as
\begin{align}
\Gamma=\left(\begin{array} {cc}
1\\
 & 1
\end{array} \right)\otimes
\left(\begin{array}{cc}
U & J/2\\
J/2 & U
\end{array}\right). \nonumber
\end{align}

\begin{figure}
\includegraphics[scale=0.6,trim=15 0 0 0]{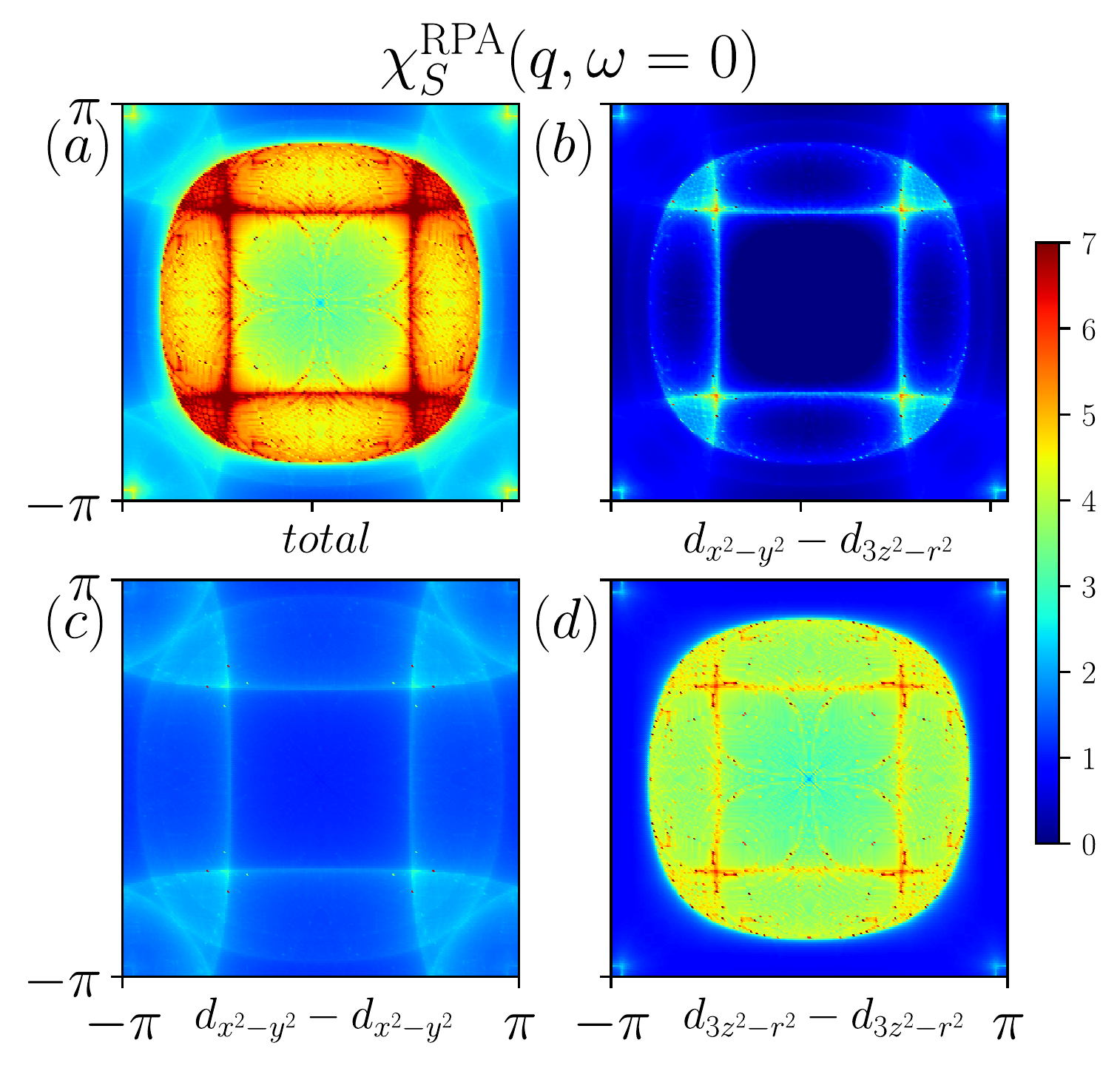}\caption{\label{fig:chi}Spin susceptibility $\chi^{st,\rm{RPA}}_S(q,\omega=0)$ of the bilayer two-orbital model. (a) The total orbital sum $\chi_S^{\mathrm{RPA}}=\sum_{st}\chi_S^{st,\mathrm{RPA}}$ which corresponds to the experimental measurable. (b-d) Orbital-resolved $\chi^{st,\rm{RPA}}$.  An amplified factor of 2 is used in (b-c).}
\end{figure}

In Fig.~\ref{fig:chi} we show the constant energy slices of $\chi_S^{\rm {RPA}}(q,\omega=0)$. Here we use $U=3, J=0.4$\ eV. $T=0$ is applied since temperature only trivially brings a broadening to the spectrum. Fig.~\ref{fig:chi}a is the total $\chi_S^{\rm RPA}=\sum_{s,t}\chi_S^{st,\rm RPA}$ corresponding to experimental measurable. As can be seen, the magnetic signal shows a ring-like enhancement.
To unveil the origin, we show in Figs.~\ref{fig:chi}b-d the orbital-resolved $\chi^{st,\rm RPA}_S$, from which we can see a dominated intra-orbital $d_{3z^2-r^2}$ scattering reflecting Fermi surface nesting of $\gamma-$pocket.  While the signal from other two channels are  weaker, consisting with the strong orbital mixing in  $\alpha,\beta-$pockets. Our  result could be further tested in the  magnetic measurement.


{\it Discussion.}$-$
The discovery of the high transition temperature superconductor  $ \mathrm{La_{3}Ni_{2}O_{7}}$  represents a major breakthrough in the field of nickletate superconductivity.  Our DFT calculations demonstrate that there are two electron pockets $\alpha,\beta$ and one hole pocket $\gamma$ on the Fermi surface, in which the $\alpha,\beta-$pockets exhibit mixing of orbitals, while the  $\gamma-$pocket  features a dominated $d_{3z^{2}-r^{2}}$ content.
In comparison to the bulk Ni-112, which has not yet demonstrated a finite $T_c$,  $ \mathrm{La_{3}Ni_{2}O_{7}}$ exhibits several distinguishing features that may be crucial to superconductivity. First, the less correlated La$-5d$ derived bands are expelled from the Fermi level, diminishing the hybridization between the Ni$-3d_{x^2-y^2}$ and La$-5d$  orbitals that impedes superconductivity. Furthermore, the site energy difference between Ni$-d_{x^2-y^2}$ and O$-p$  in $ \mathrm{La_{3}Ni_{2}O_{7}}$ is estimated as $ \Delta \equiv \epsilon _d - \epsilon _p$ = 3.88\ eV. This value is smaller than that of $ \mathrm{RNiO_{2}}$ (4.4 eV) \cite{botana2020}, which  could potentially contribute to the high $T_c$ in  $ \mathrm{La_{3}Ni_{2}O_{7}}$ in the context of pairing based on Zhang-Rice singlet state \cite{zhang1988}. The inclusion of the $d_{3z^2-r^2}$ near Fermi level in  $ \mathrm{La_{3}Ni_{2}O_{7}}$, however, may have complex implications for superconductivity. The large density of state of $d_{3z^2-r^2}$ orbital can provide new phase space for the potential pairing of electrons \cite{hjp12}. However, the presence of multiple orbitals on Fermi level may also lead to competition between pairings with different symmetries, such as the competition between $d_{x^2-y^2}$ and $s_{\pm}$ wave pairing. Regarding the filling factors of the relevant orbitals, we note that in the case of  a $d^{7.5}$ configuration of Ni, if $d_{3z^2-r^2}$ is considered to roughly have the same occupation number of $d_{x^2-y^2}$, then  both orbitals have about $0.75$  electrons per site, corresponding to 25\% hole-doping. It is notable that the oxygen-deficient in realistic materials may effectively reduce the hole doping level in  e$_g$ orbitals of Ni$-3d$,  resulting enhanced superconductivity. Finally,  we acknowledge that the $d_{3z^2-r^2}$ orbitals exhibit much weaker hybridization with in-plane oxygen compared to its $d_{x^2-y^2}$ counterpart, which necessitates in-depth investigations into its strong interaction effects and its influence on superconductivity. The question about the role of the electron-phonon coupling, which becomes specifically important since the superconductivity  in  $ \mathrm{La_{3}Ni_{2}O_{7}}$  is found under pressure,  should also be clarified in future studies.

{\it Conclusion.}$-$ In conclusion, we have introduced a minimal bilayer two-orbital model for the Ruddlesden-Popper bilayer La$_3$Ni$_2$O$_7$ under pressure. The tight-binding parameters are obtained by Wannier downfolding of the DFT calculations, which reproduce the band structure and Fermi surface well.  The spin susceptibility is studied  using the RPA method, which shows that the magnetic signal
majorly comes  from $d_{3z^2-r^2}$. This model provides important means to the study of electronic, magnetic, orbital, and superconducting properties of the material under pressure.


We thank the useful discussions with Guang-Ming Zhang and Biao Lv. Work at Sun Yat-Sen University was supported by the National Key Research and Development Program of China (Grants No. 2022YFA1402802, 2018YFA0306001), the National Natural Science Foundation of China (Grants No. 92165204, No.12174454, No. 11974432, No.12274472),  the Guangdong Basic and Applied Basic Research Foundation (Grants No. 2022A1515011618, No. 2021B1515120015), Guangdong Provincial Key Laboratory of Magnetoelectric Physics and Devices (Grant No. 2022B1212010008), Shenzhen International Quantum Academy (Grant No. SIQA202102), and Leading Talent Program of Guangdong Special Projects (201626003). 
%
\end{document}